\documentclass[conference]{IEEEtran}
\IEEEoverridecommandlockouts
\usepackage{cite}
\usepackage{amsmath,amssymb,amsfonts}
\usepackage{algorithmic}
\usepackage{lipsum,adjustbox}
\usepackage{verbatim}
\usepackage{makecell}
\usepackage{graphics}
\usepackage{stackengine}
\usepackage{pgfplots}
\pgfplotsset{width=7cm,compat=1.8}
\usepackage{multirow}
\usepackage{soul}
\usepackage{amssymb}
\usepackage{amsmath}
\usepackage[caption=false,font=footnotesize]{subfig}
\usepackage{algorithm}
\usepackage{algorithmic}
\usepackage{pgfplots}
\pgfplotsset{compat=1.18}
\usepackage{graphicx}
\usepackage{textcomp} 
\usepackage{booktabs}
\usepackage{stfloats}
\usepackage{scalefnt}
\usepackage{mathrsfs}       
\DeclareMathAlphabet{\pazocal}{OMS}{zplm}{m}{n}
\usepackage[nolist]{acronym}
    
 \usepackage{bm}

\newcommand{\compl}{\mathbb{C}}         

\newcommand{\ma}  [1]{ \bm{#1} } 

\newcommand{\set} [1]{{\mathcal {#1}}} 
\newcommand{\Kon} {\set{K}_{\text{on}}} 
\newcommand{\Kp} {\set{K}_{\text{p}}} 
\newcommand{\Kd} {\set{K}_{\text{d}}} 

\begin{acronym}
    \acro{C-ITS}{cooperative intelligent transportation system}
    \acro{RBF}{radial basis function}
    \acro{ANN}{artificial neural network}
\acro{GRU}{gated recurrent unit}
\acro{RNN}{Recurrent neural network}
\acro{Bi-RNN}{bi-directional recurrent neural network}
\acro{Bi}{bi-directional}
\acro{FNN}{Feed-forward Neural Network}
    \acro{CNN}{convolutional neural network}
    \acro{TS-ChannelNet}{Temporal spectral ChannelNet}
    \acro{TDR}{transmission data rate}
    \acro{LSTM}{long short-term memory}
    \acro{ALS}{Accurate LS}
    \acro{SLS}{simple LS}
    \acro{LRP}{Layer-wise Relevance Propagation}
    \acro{XAI}{explainable artificial intelligence}
    \acro{ChannelNet}{channel network}
    \acro{ADD-TT}{average decision-directed with time truncation}
    \acro{WI}{weighted interpolation}
    \acro{DD}{decision-directed}
    \acro{SR-ConvLSTM}{super-resolution convolutional long short-term memory}
    \acro{SR-CNN}{super resolution CNN}
    \acro{DN-CNN}{denoising CNN}
    \acro{V2V}{vehicle-to-vehicle}
    \acro{V2I}{vehicle-to-infrastructure}
    \acro{VTV}{vehicle-to-vehicle}
    \acro{RTV}{roadside-to-vehicle}
    \acro{DPA}{data-pilot aided}
    \acro{STA}{spectral temporal averaging}
    \acro{M-STA}{modified STA}
    \acro{LMMSE}{linear minimum mean-squared error}
    \acro{M-CDP}{modified CDP}
    \acro{M-TRFI}{modified TRFI}
    \acro{SNR}{signal-to-noise ratio}
    \acro{CDP}{Constructed data pilots}
    \acro{DFT}{discrete Fourier transform}
    \acro{T-DFT}{truncated discrete Fourier transform}
    \acro{TA-MDFT}{temporal averaging M-DFT}
    \acro{TA-DFT}{TA-DFT}
    \acro{STS}{short training symbols}
    \acro{LTS}{long training symbols}
    \acro{SS}{signal symbol}
    \acro{TDL}{tapped delay line}
    \acro{TRFI}{time domain reliable test frequency domain interpolation}
    \acro{SBS}{symbol-by-symbol}
    \acro{FBF}{frame-by-frame}
    \acro{MMSE-VP}{minimum mean square error using virtual pilots}
    \acro{DL}{deep learning}
    \acro{AE-DNN}{auto-encoder deep neural network}
    \acro{DNN}{deep neural networks} 
    \acro{ISI}{inter-symbol-interference}
    \acro{OFDM}{Orthogonal Frequency-Division Multiplexing}
    \acro{LS}{least square}
    \acro{RS}{reliable subcarriers}
    \acro{URS}{unreliable subcarriers}
    \acro{SRNN}{Simple RNN}
    \acro{MMSE}{minimum mean squared error}
    \acro{SoA}{state of the art}
    \acro{NMSE}{normalized mean-squared error}
    \acro{AWGN}{additive white Gaussian noise}
	\acro{BER}{Bit Error Rate}
	\acro{RSU}{road side unit}
	\acro{AE}{auto-encoder}
    \acro{CP}{cyclic prefix}
\acro{PLCP}{physical layer convergence protocol}
\acro{MSE}{mean squared error}
\end{acronym}

\usepackage[utf8]{inputenc}
\usepackage[english]{babel}
\usepackage{amsthm}
\usepackage{commath}

\newcommand{\Lb}{\pazocal{L}}

\usepackage{cleveref}
\crefname{lemma}{Lemma}{Lemmas}
\usepackage{flushend}

\begin{document}

\title{X-RACE: XAI-assisted Recurrent neural network Attribution for Channel Estimation
}

\author{\IEEEauthorblockA{Abdul Karim Gizzini\IEEEauthorrefmark{1},
Yahia Medjahdi\IEEEauthorrefmark{4}
}

\IEEEauthorblockA{\IEEEauthorrefmark{1}University of Paris-Est Créteil (UPEC), LISSI/TincNET, F-94400, Vitry-sur-Seine, France.\\
\IEEEauthorrefmark{4} IMT Nord Europe, Institut Mines T\'el\'ecom, Centre for Digital Systems, F-59653 Villeneuve d’Ascq, France. \\
Email: abdul-karim.gizzini@u-pec.fr, yahia.medjahdi@imt-nord-europe.fr}}

\maketitle

\begin{abstract}

Deep learning models, notably Long Short-Term Memory (LSTM), have demonstrated promising performance in channel estimation for high-mobility vehicular environments. However, their black-box nature and architectural overhead limit trustworthiness and efficiency. Classical explainable AI (XAI) methods rely on costly iterative processes, offering only input-level filtering without addressing architectural fine-tuning. To overcome these limitations, this paper proposes the XAI-assisted Recurrent neural network Attribution for Channel Estimation (X-RACE) framework. X-RACE uses a low-complexity, one-shot dual-optimization strategy to simultaneously evaluate and prune irrelevant input subcarriers and internal hidden units. Furthermore, we propose novel temporal XAI metrics: Saturation Time, Importance Drift, and Relevance Contrast to characterize the LSTM's learning dynamics and memory convergence. Extensive simulations demonstrate that X-RACE reduces inference complexity by at least $44.1\%$ while improving or preserving Bit Error Rate (BER) performance, outperforming classical XAI schemes.

\end{abstract}


\section{Introduction} \label{introduction}

Artificial intelligence (AI) serves as a foundational pillar for next-generation 6G networks~\cite{11063291, 11585835}, where Deep Learning (DL) models are employed in the physical layer to support the high data rate and latency demands of mission-critical applications~\cite{10929033, 11574706}. Reliable receiver operation depends on accurate channel estimation, which is challenging in high-mobility vehicular environments due to severely doubly-selective channels. To bypass the performance degradation of classical estimation schemes, DL-based post-processing models are widely adopted~\cite{10540188}. While early implementations employ memoryless Feedforward Neural Networks (FNNs)~\cite{gizzini2025explainable}, they ignore underlying temporal correlations. Conversely, Long Short-Term Memory (LSTM) networks model long-term temporal dependencies, making them well-suited for high-mobility tracking~\cite{9954409}. LSTM addresses double selectivity by combining an input feature vector capturing spectral snapshots with internal recurrent states tracking Doppler impact over time.
Despite their superior accuracy, standard LSTMs operate as black boxes, making it difficult to understand how their dense internal parameters actually learn time-varying channel dynamics. This lack of transparency undermines trustworthiness in safety-critical applications~\cite{10854503, 10620685}. Furthermore, standard LSTMs often use excessive internal hidden units, leading to unnecessary computational overhead. Classical eXplainable Artificial Intelligence (XAI) schemes, such as Local Interpretable Model-agnostic Explanations (LIME)~\cite{ribeiro2016should} and SHapley Additive exPlanations (SHAP)~\cite{lundberg2017unified}, offer input-level feature filtering but fail to prune the LSTM architecture. Additionally, these iterative methods induce prohibitively high computational complexity. Alternatively, the uniform input downsampling proposed in~\cite{9954409} reduces the architecture empirically rather than a context-aware mechanism adapted to varying channel conditions.

To address these challenges, and inspired by~\cite{gizzini2025explainable}, this paper proposes the XAI-assisted Recurrent neural network Attribution for Channel Estimation (X-RACE) framework. X-RACE utilizes a low-complexity, one-shot dual-optimization strategy that simultaneously introduces dynamic perturbation noise into a pre-trained LSTM's inputs and internal memory state. This joint attribution isolates specific relevant subcarriers and internal hidden units that contribute to channel tracking. Furthermore, to explicitly capture the recursive nature of the LSTM, X-RACE incorporates novel XAI-based temporal metrics to evaluate model convergence and relevance stability. The main contributions of this work are summarized as follows:

\begin{itemize}
\item Propose X-RACE, a dual-optimization input and architectural attribution framework specifically designed for recurrent channel estimators in doubly-selective channels.

\item Introduce sliding-window XAI-based metrics to measure the learning dynamics and memory convergence of the employed LSTM model.

\item Demonstrate via extensive simulations that X-RACE's input-architecture optimization prunes irrelevant inputs and internal hidden units, outperforming classical XAI schemes by substantially reducing computational complexity while preserving \ac{BER} performance.
\end{itemize}

The remainder of this paper is organized as follows: Section II presents the LSTM-based channel estimation. Section III details the proposed X-RACE framework. Section IV analyzes performance and computational complexity, and Section V concludes the paper.
\section{LSTM-based Channel Estimation} \label{system_model}

We consider an {\ac{OFDM}} system. $K_{\text{on}} = K_{p} + K_{d}$ represents the active subcarriers. $K_{p}$ and $K_{d}$ refer to the allocated pilot and data subcarriers, respectively. The $i$-th received {\ac{OFDM}} symbol $\tilde{\ma{y}}_i \in \compl^{K_{\text{on}} \times 1}$ can be expressed as:

\begin{equation}
   \tilde{\ma{y}}_i[k] = \tilde{\ma{h}}_{i}[k] \tilde{\ma{x}}_{i}[k] + \tilde{\ma{e}}_{i}[k] +\tilde{\ma{v}}_{i}[k],\quad k \in \Kon 
\label{eq: xK}
\end{equation}
where $\tilde{\ma{x}}_{i} \in \compl^{K_{\text{on}} \times 1}$ and $\tilde{\ma{h}}_{i} \in \compl^{K_{\text{on}} \times 1}$ denote the $i$-th transmitted {\ac{OFDM}} symbol and its respective time variant frequency-domain channel response. Moreover, $\tilde{\ma{v}}_{i} \in \compl^{K_{\text{on}} \times 1}$ and $\tilde{{\ma{e}}}_{i} \in \compl^{K_{\text{on}} \times 1}$ represent the \ac{AWGN} and the  Doppler-induced inter-carrier interference. 

LSTM models effectively track doubly-selective channels across OFDM frames. This is achieved by employing interacting gates to dynamically control the flow of information at each time-step $i$, according to four sequential stages:

\subsubsection{Long-Term Memory Gating}  The forget gate selectively discards irrelevant historical components from the long-term memory, outputting a scaling vector via the sigmoid activation function ${\sigma}(\cdot)$, such that:

\begin{equation}
{\ma{f}}_{i} = {\sigma} (\ma{W}_{f}\hat{{\ma{h}}}^{\prime}_{{i}} + \ma{U}_{f}{\ma{s}}_{i-1} + {\ma{b}}_{f}).
\label{eq:lstm_fg}
\end{equation}

$\ma{W}_{f} \in \mathbb{R}^{S \times 2K_{\text{on}}}$, $\ma{U}_{f} \in \mathbb{R}^{S \times S}$, and ${\ma{b}}_{f} \in \mathbb{R}^{S \times 1}$ are the forget gate input weights, recurrent weights, and bias, respectively. $S$ is the hidden state dimension, $\hat{\ma{h}}^{\prime}_{i} \in \mathbb{R}^{2K_{\text{on}} \times 1}$ is the current estimated channel, and ${\ma{s}}_{i-1} \in \mathbb{R}^{S \times 1}$ represents the previous short-term hidden state.

\subsubsection{Current Input Gating and Candidate Generation} 
Concurrently, the input gate evaluates the current channel estimate to determine the update magnitude for each memory coordinate, while a candidate cell state $\tilde{{\ma{c}}}_{i}$ models potential adjustments via a $\tanh$ activation, such that:

\begin{equation}
{{\ma{g}}_{i}} = {\sigma} (\ma{W}_{g}\hat{\ma{h}}^{\prime}_{i} + \ma{U}_{g}{\ma{s}}_{i-1} + {\ma{b}}_{g}),
\label{eq:lstm_ing}
\end{equation}
\begin{equation}
{\tilde{{\ma{c}}}}_{i} = \tanh (\ma{W}_{c}\hat{\ma{h}}^{\prime}_{i} + \ma{U}_{c}{\ma{s}}_{i-1} + {\ma{b}}_{c}),
\label{eq:lstm_incg}
\end{equation}
where $\ma{W}_{g}, \ma{W}_{c} \in \mathbb{R}^{S \times 2K_{\text{on}}}$, $\ma{U}_g, \ma{U}_c \in \mathbb{R}^{S \times S}$, and ${\ma{b}}_g, {\ma{b}}_c \in \mathbb{R}^{S \times 1}$ are the input weights, recurrent weights, and biases for the input and candidate gates, respectively.

\subsubsection{Long-Term Cell State Synchronization}

The long-term cell state  $\ma{c}_{i} \in \mathbb{R}^{S \times 1}$ is updated by blending past and present information, using an element-wise Hadamard product $\odot$ to apply the forget and input scales:

\begin{equation}
{{{\ma{c}}}}_{i} = {\ma{f}}_{i} \odot {\ma{c}}_{i-1} + {\ma{g}}_{i} \odot {\tilde{{\ma{c}}}}_{i}.
\label{eq:lstm_cell_state}
\end{equation}

\subsubsection{Hidden State and Output Generation} 

Finally, the updated hidden state ${\ma{s}}_{i}$ is constructed to provide the instantaneous channel tracking output. The output gate filters the synchronized cell state memory, such that:

\begin{equation}
{\ma{o}}_{i} = {\sigma} (\ma{W}_{o}\hat{\ma{h}}^{\prime}_{i} + \ma{U}_{o}{\ma{s}}_{i-1} + \bar{\ma{b}}_{o}),
\label{eq:lstm_og}
\end{equation}
\begin{equation}
{{{\ma{s}}}}_{i} = {\ma{o}}_{i} \odot \tanh(\ma{c}_{i}).
\label{eq:lstm_hidden_state}
\end{equation}

Finally, an FNN layer is employed to extract the final LSTM-based channel estimate $\hat{{\ma{h}}}_{\text{LSTM}_{i}} \in \mathbb{R}^{2K_{\text{on}} \times 1}$ from the current hidden state ${\ma{s}}_{i}$. Despite their tracking capabilities, equations {\eqref{eq:lstm_fg}}-{\eqref{eq:lstm_hidden_state}} form an opaque black box. The dense internal parameter interactions make it difficult to verify how the LSTM allocates its internal hidden units to adapt to doubly-selective channels. This transparency limitation necessitates a diagnostic framework for joint input and internal hidden units filtering without altering the pre-trained model performance, thereby motivating the proposed X-RACE framework.
\section{Proposed X-RACE Framework} \label{sec:proposed_x_race}

\begin{figure*}
    \centering
\includegraphics[width=1\linewidth]{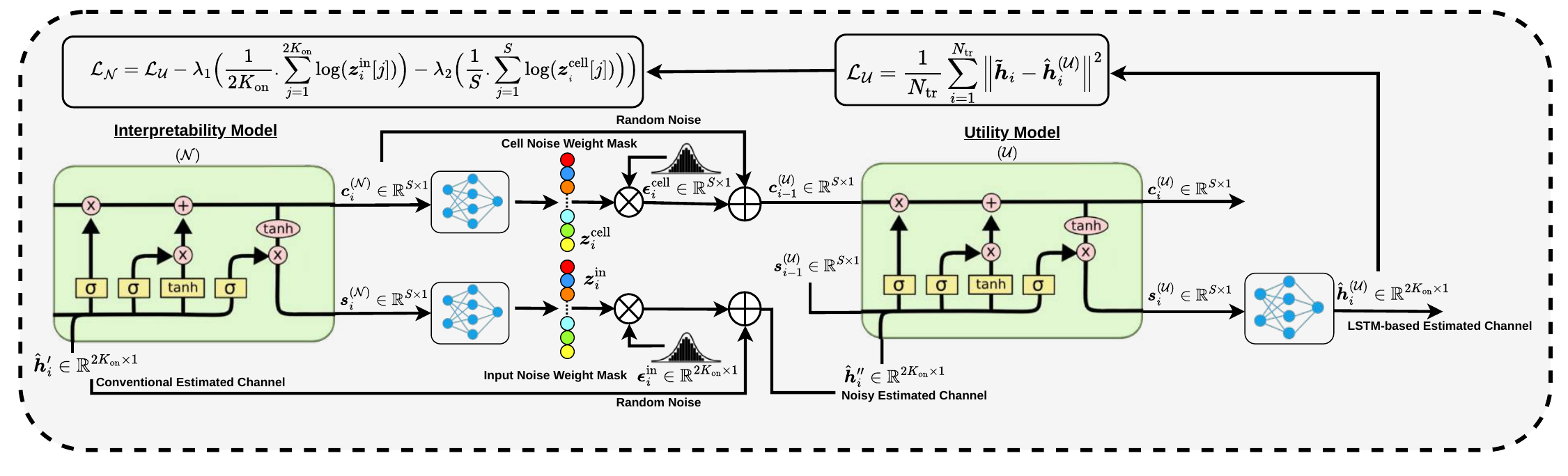}
    \caption{Block Diagram of the proposed X-RACE framework.}
    \label{fig:xrace_bd}
\end{figure*}

Let $\mathcal{U}$ denote the pre-trained LSTM black-box utility model operating as the primary channel estimator, parameterized by $S$ hidden state dimension. To evaluate the time-varying sensitivity of $\mathcal{U}$, the proposed X-RACE framework proceeds according to the following steps, as shown in Figure 1:

\subsubsection{Joint Noise Mask Generation}

First, an interpretability noise LSTM model $\mathcal{N}$ is introduced to track the sequential channel inputs and generate a time-varying, subcarrier-specific perturbation mask $\ma{z}^{\text{in}}_{i} \in [0, 1]^{2K_{\text{on}} \times 1}$, and a cell state sensitivity mask $\ma{z}^{\text{cell}}_{i} \in [0, 1]^{S \times 1}$. In this context, the $i$-th conventionally estimated channel\footnote{The real and imaginary components of the $K_{\text{on}}$ active subcarriers are stacked vertically.} $\hat{\ma{h}}^{\prime}_{i} \in \mathbb{R}^{2K_{\text{on}} \times 1}$  is fed to the $\mathcal{N}$ model to generate both perturbation masks, such that:

\begin{equation}
\ma{z}^{\text{in}}_{i}, \ma{z}^{\text{cell}}_{i}, \ma{s}_{i}^{(\mathcal{N})}, \ma{c}_{i}^{(\mathcal{N})} = \mathcal{N}\left(\hat{\ma{h}}^{\prime}_{i}, \ma{s}_{i-1}^{(\mathcal{N})}, \mathbf{c}_{i-1}^{(\mathcal{N})}; S\right).
\end{equation}

$\ma{s}_{i}^{(\mathcal{N})}$ and $\ma{c}_{i}^{(\mathcal{N})}$ represent the $i$-th hidden and cell state vectors of the interpretability LSTM model, respectively. The elements of the generated perturbation masks are bounded within the range $[0, 1]$ via a sigmoid activation.

After that, the generated perturbation masks $\ma{z}^{\text{in}}_{i}$, and $\ma{z}^{\text{cell}}_{i}$ are multiplied by random noise vectors $\epsilon^{\text{in}}_{i} \sim  \mathcal{N} (0,1)$ and $\epsilon^{\text{cell}}_{i} \sim  \mathcal{N} (0,1)$ sampled from the standard normal distribution, respectively. The weight noise vectors are then added to the conventional estimated channel and the cell state, such that:

\begin{equation} 
\hat{\ma{h}}^{\prime\prime}_{i} = \hat{\ma{h}}^{\prime}_{i} + \ma{z}^{\text{in}}_{i} \odot \boldsymbol{\epsilon}_{i}^{\text{in}}, ~~~~
{\ma{c}}_{i-1}^{(\mathcal{U})} = \ma{c}_{i-1}^{(\mathcal{U})} + \ma{z}^{\text{cell}}_{i} \odot \boldsymbol{\epsilon}_{i}^{\text{cell}}.
\end{equation}

\subsubsection{Utility Model Tracking with Perturbed States} 

The objective of (8) and (9) is to perturb the input and long-term memory of the pre-trained $\mathcal{U}$ model, simultaneously.
The frozen utility model $\mathcal{U}$ processes the corrupted input sequence using the corrupted cell state memory, such that:

\begin{equation}
\hat{\ma{h}}^{(\mathcal{U})}_{i}, \ma{s}_{i}^{(\mathcal{U})}, \ma{c}_{i}^{(\mathcal{U})} = U\left(\hat{\ma{h}}^{\prime\prime}_{i}, \ma{s}_{i-1}^{(\mathcal{U})}, {\ma{c}}_{i-1}^{(\mathcal{U})}; S\right).
\end{equation}

\subsubsection{Customized Joint Optimization Loss Function}

The interpretability model $\mathcal{N}$ is trained to quantify the relevance of both input subcarriers and internal hidden units. This is performed by optimizing the learnable $\ma{z}^{\text{in}}_{i}$, and $\ma{z}^{\text{cell}}_{i}$ while minimizing the loss function $\mathcal{L}_{\mathcal{U}}$ of the pre-trained $\mathcal{U}$ model, such that:

\begin{equation}
\mathcal{L}_{\mathcal{U}} = \frac{1}{N_{\text{tr}}} \sum_{i=1}^{N_{\text{tr}}} \left( \tilde{\ma{h}}_{i} - \hat{\ma{h}}^{(\mathcal{U})}_{i} \right)^2.
\end{equation}

The input and cell learnable noise masking growth terms $\ma{z}^{\text{in}}_{i}$, and $\ma{z}^{\text{cell}}_{i}$ can be expressed as follows: 

\begin{equation}
\small
\mathcal{L}_{\text{input}} = \frac{1}{2K_{\text{on}}} \sum_{j=1}^{2K_{\text{on}}} \log\left(\ma{z}^{\text{in}}_{i}[j]\right), ~~\mathcal{L}_{\text{cell}} = \frac{1}{S} \sum_{m=1}^{S} \log\left(\ma{z}^{\text{cell}}_{i}[m]\right).
\end{equation}

The training of the $\mathcal{N}$ model aims to minimize a customized joint objective function $\mathcal{L}_{\mathcal{N}}$ that balances tracking accuracy of $\mathcal{L}_{\mathcal{U}}$ while maximizing $\ma{z}^{\text{in}}_{i}$ and $\ma{z}^{\text{cell}}_{i}$, weighted by hyperparameters $\lambda_1$ and $\lambda_2$, respectively. Therefore, the objective is to push $\ma{z}^{\text{in}}_{i}$ and $\ma{z}^{\text{cell}}_{i}$ toward unity for irrelevant inputs and internal hidden units, such that:

\begin{equation}\mathcal{L}_{\mathcal{N}} = \mathcal{L}_{\mathcal{U}} - \lambda_1 \mathcal{L}_{\text{input}} - \lambda_2 \mathcal{L}_{\text{cell}}.
\end{equation}

Building upon the learnable $\ma{z}^{\text{in}}_{i}$ and $\ma{z}^{\text{cell}}_{i}$, the X-RACE double optimization problem aims to minimize the Mean Squared Error (MSE) between the true channel, $\tilde{\ma{h}}_{i}$, and the optimized LSTM-based channel estimate, such that:

\begin{equation}
\small
\begin{aligned}
\min_{\tau, P} \quad &  \Lb_{\mathcal{U}^{*}} = \frac{1}{N_{\text{tr}}}. \sum_{i = 1}^{N_{tr}} \left( {\tilde{\ma{h}}}_i - {\mathcal{U}}^{*} \Big( \hat{{\ma{h}}}_{{i}} \odot \ma{m}^{*}_{\text{in}} (\tau) ; \ma{c}_{i} \odot \ma{m}^{*}_{\text{cell}}(P)\Big)\right)^{2},\\
\textrm{s.t.} \quad &  \ma{m}^{*}_{\text{in}} = \mathbb{I} \left( \bar{\ma{r}}_{\text{in}} \geq \tau \right), \quad \tau \in \mathcal{T}, \\
& \ma{m}^{*}_{\text{cell}} = \mathbb{I} \left( \bar{\ma{r}}_{\text{cell}} \geq \text{Percentile}(\bar{\ma{r}}_{\text{cell}}, P) \right), P \in \mathcal{P}, \\
& \text{BER}(\mathcal{U}^{*}) \le \text{BER}(\mathcal{U}).
\end{aligned}
\label{eq:threshold_fine_tuning}
\end{equation}

$N_{\text{tr}}$ denotes the number of training samples, and $\mathbb{I}(\cdot)$ is the indicator function. The terms $\bar{\ma{r}}_{\text{in}} =  1 - \mathbb{E} [\ma{z}^{\text{in}}]$ and $\bar{\ma{r}}_{\text{cell}} =  1 - \mathbb{E} [\ma{z}^{\text{cell}}]$ represent the relevance scores assigned to the $\mathcal{U}$ model inputs and internal hidden units, respectively. The relevance is the complement of the induced noise such that higher values reflect greater importance. Furthermore, $\tau$ and $P$ denote the optimal input relevance threshold and architectural pruning percentile, defined such that $\mathcal{T} = \{ \tau \in \mathbb{R} \mid \tau_{\text{min}} \leq \tau \leq \tau_{\text{max}}, \tau = \tau_{\text{min}} + n_{\tau} \Delta\tau\}$ and $\mathcal{P} = \{ P \in \mathbb{Z} \mid 0 \leq P \leq P_{\text{max}}, P = n_P \Delta P\}$, with step sizes $\Delta\tau$ and $\Delta P$. Finally, the constraint ensures that the average \ac{BER} using the pruned LSTM model $\mathcal{U}^{*}$ does not exceed that of the full baseline $\mathcal{U}$.

\subsubsection{Proposed XAI-Assisted Temporal Metrics} To characterize the temporal dynamics of the LSTM memory adaptation, we propose three primary XAI-assisted metrics: 

\begin{itemize}
    \item Saturation Time ($T_{\text{sat}}$): The earliest temporal index where the LSTM achieves a stable noise update over a window $W$ over $I$ symbols per frame. To mitigate fluctuations, a moving average filter of length $L$ is first applied to smooth the averaged $\ma{z}^{\text{in}}_{i}$, such as:

    \begin{equation}
    \tilde{\ma{z}}_{i}[k] = \frac{1}{L} \sum_{m=-\lfloor L/2 \rfloor}^{\lfloor L/2 \rfloor} \mathbb{E}\big[\ma{z}^{\text{in}}_{i+m}[k]\big].
    \end{equation}
      
    The saturation time $T_{\text{sat}}$ is then defined as:

    \begin{equation}
    \small
    T_{\text{sat}} [k] = \min_{ i \in [1, I - W]} : \max_{\tau \in [i, i+W-1]} \left| \tilde{\ma{z}}_{ \tau+1} [k] - \tilde{\ma{z}}_{\tau} [k]\right| < \epsilon_{\text{local}} [k],
    \end{equation}
    where $\epsilon_{\text{local}}$ sensitivity is dynamically scaled by the subcarrier's temporal dynamic range to ensure identification robustness:
    
    \begin{equation}
    \epsilon_{\text{local}}[k] = \epsilon_{\text{base}} \cdot \left( \max_{i \in [1, I]} { \tilde{\ma{z}}_{i}[k] } - \min_{i \in [1, I]} { \tilde{\ma{z}}_{i}[k] } + 1 \right).
    \end{equation}

    \item Importance Drift ($\bar{\mathcal{D}}_{\Phi}$):  Measures the average magnitude of the adaptation effort for data and pilot subcarriers, defined as the mean absolute displacement from the initial induced noise at $i=1$ to the noise at $T_{\text{sat}}$, such that:
    
    \begin{equation}
    \bar{\mathcal{D}}_{\Phi} = \frac{1}{|\Phi|} \sum_{k = 1}^{|\Phi|} \left| \tilde{\ma{z}}_{T_{\text{sat}}} [k] - \tilde{\ma{z}}_{1}[k] \right|, ~~ \Phi \in \{ \Kd,~\Kp\}.
    \end{equation}
    
    \item Relevance Contrast ($\Delta \bar{\mathcal{R}}$): Quantifies the steady-state discriminative trust gap between data and pilot subcarriers expressed as follows:
    
    \begin{equation}
    \Delta \bar{\mathcal{R}} = \frac{1}{|\Kd|} \sum_{k = 1}^{K_{d}} \tilde{\ma{z}}_{T_{\text{sat}}} [k] - \frac{1}{|\Kp|} \sum_{k = 1}^{K_{p}} \tilde{\ma{z}}_{T_{\text{sat}}} [k].
    \end{equation}
\end{itemize}
\section{Simulation Results} \label{simulation_results}

The X-RACE framework\footnote{A direct BER comparison with the XAI-CHEST framework in [6] is omitted. Fundamental differences in the training dataset shaping cause FNNs to undergo more gradient updates per epoch than LSTMs under an identical batch size, rendering a direct BER comparison mathematically unfair.} is evaluated using the DPA-LSTM-NN estimator \cite{9954409}. Uniform input downsampling~\cite{9954409}, SHAP~\cite{lundberg2017unified}, and LIME~\cite{ribeiro2016should} are employed as benchmark XAI schemes.
The IEEE 802.11p standard is used with $K_{p} = 4$, $K_{d} = 48$, $K_{n} = 12$, and $I = 50$ OFDM symbols per frame, with a total symbol duration of $T_{\text{OFDM}} = 8~\mu\text{s}$. The employed channel models are the high-mobility ($F_d=1000\text{ Hz}$) VTV-EX low-frequency selective (LF) and VTV-SDWW high-frequency selective (HF) scenarios \cite{r19}. The model is trained on $10^5$ frames using an 80\%/20\% train-test split. Optimization is performed using the Adam optimizer for 500 epochs with a batch size of 128. For the proposed XAI metrics, we set $L = 3$ and $W = 5$ symbols ($10\%$ of the frame) to prevent premature convergence on local fluctuations. We note that $T_{\text{sat}}$ is inversely proportional to the $\epsilon_{\text{base}}$ sensitivity, as tighter tolerances naturally delay convergence. Because the overall conclusions remain consistent regardless of the specific value of $\epsilon_{\text{base}}$, we set $\epsilon_{\text{base}} = 10^{-3}$ to balance early convergence and long-term saturation. Finally, the performance analysis is structured across four criteria: (\textit{i}) LSTM learning dynamics over time selectivity, (\textit{ii}) the impact of frequency selectivity, (\textit{iii}) the impact of modulation order, and (\textit{iv}) a comprehensive XAI and inference computational complexity analysis.

\begin{figure}
    \centering
\includegraphics[width=\linewidth]{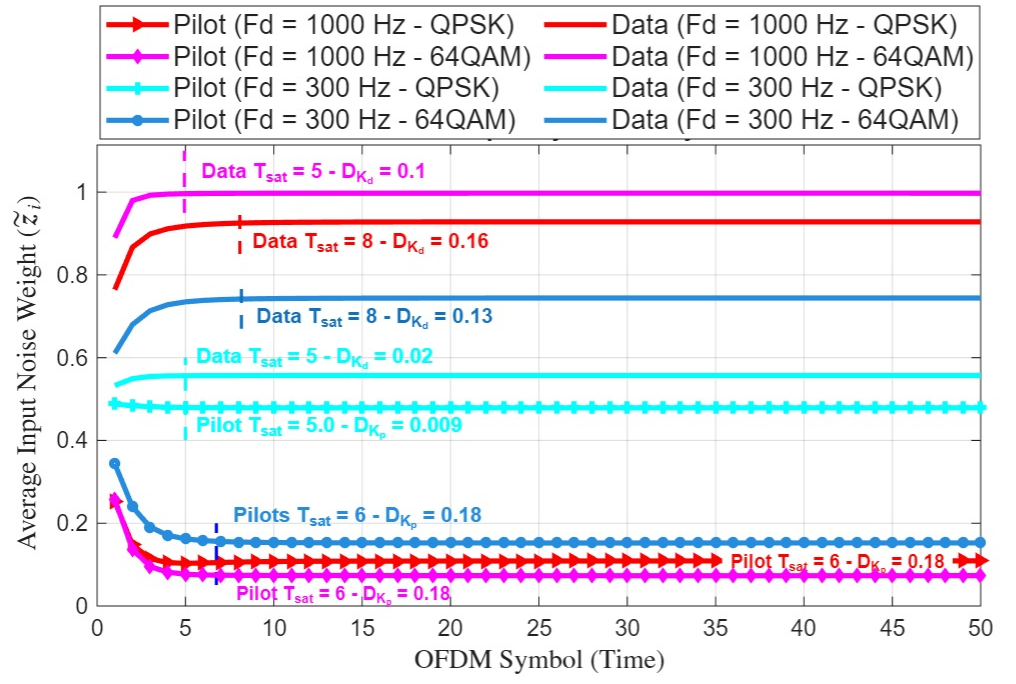}
    \caption{Temporal evolution of pilot and data averaged noise weights under varying $F_d$ and modulations, annotated with the proposed XAI metrics.}
    \label{fig:placeholder1}
\end{figure}

\subsection{LSTM Learning Dynamics: Time Selectivity Analysis}

Figure 2 evaluates $T_{\text{sat}}$, $\bar{\mathcal{D}}_{\Phi}$, and $\Delta \bar{\mathcal{R}}$ across different Doppler frequencies ($F_d$) and modulations in the LF scenario. Under low Doppler and QPSK modulation, data and pilot subcarriers show similar convergence ($T_{\text{sat}} = 5 T_{\text{OFDM}}$, $ \tilde{\ma{z}}_{i} \approx 0.5$) due to high temporal correlation. Shifting to 64QAM widens $\Delta \bar{\mathcal{R}}$, as  the LSTM prioritizes pilots ($T_{\text{sat}} = 6T_{\text{OFDM}}$, $\tilde{\ma{z}}_{i} \approx 0.19$) while increasingly masking data subcarriers. 
At a high Doppler, channel decorrelation forces the LSTM to heavily rely on pilots, as evidenced by reduced $\tilde{\ma{z}}_{i}$ and aggressive data subcarrier filtering, with an increased $T_{\text{sat}}$ due to relevance uncertainty. Furthermore, higher $F_d$ significantly increases $\bar{\mathcal{D}}_{\Phi}$ and $\Delta \bar{\mathcal{R}}$. Thus, the proposed metrics illustrate that the LSTM dynamically adapts its behavior based on its context awareness of the employed scenario.

\begin{figure*}[t]
    \centering
    \subfloat[LF - QPSK.]{\includegraphics[width=0.25\textwidth]{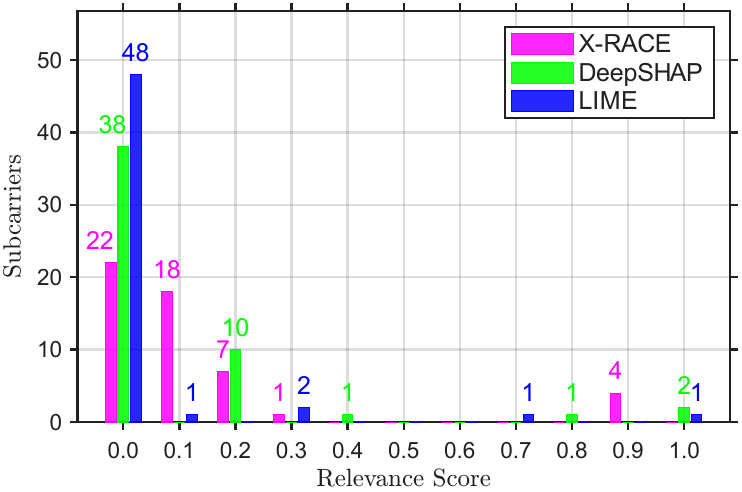}}%
    \hfill
    \subfloat[HF - QPSK.]{\includegraphics[width=0.25\textwidth]{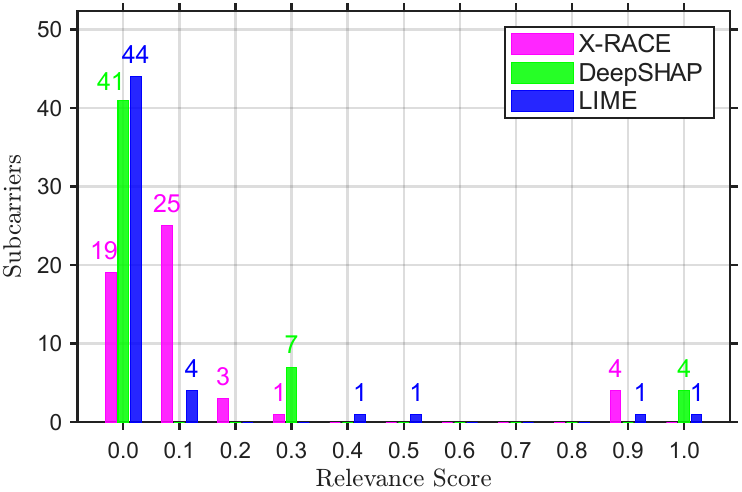}}%
    \hfill
    \subfloat[LF - 64QAM.]{\includegraphics[width=0.25\textwidth]{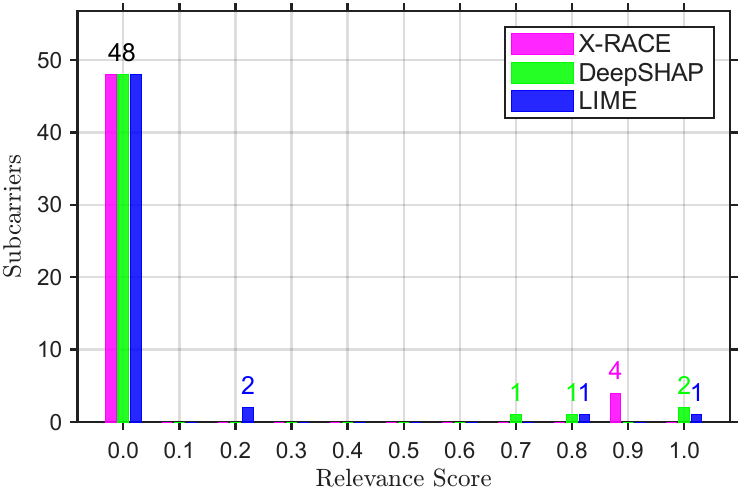}} 
    \hfill
    \subfloat[HF - 64QAM.]{\includegraphics[width=0.25\textwidth]{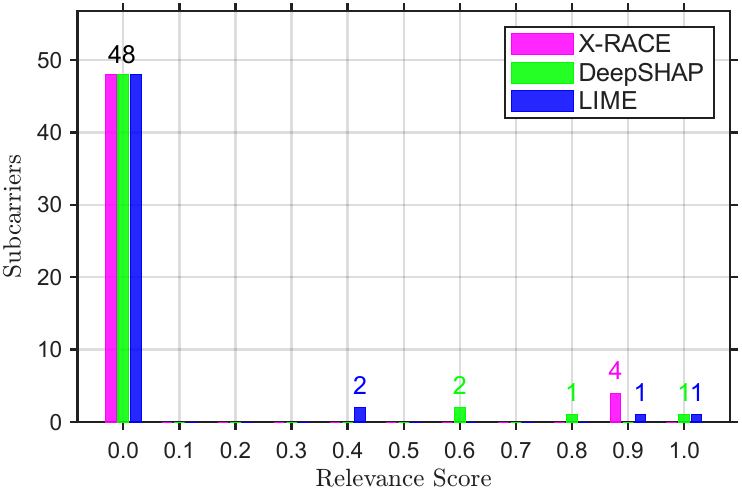}} \\
    \subfloat[LF - QPSK.]{\includegraphics[width=0.25\textwidth]{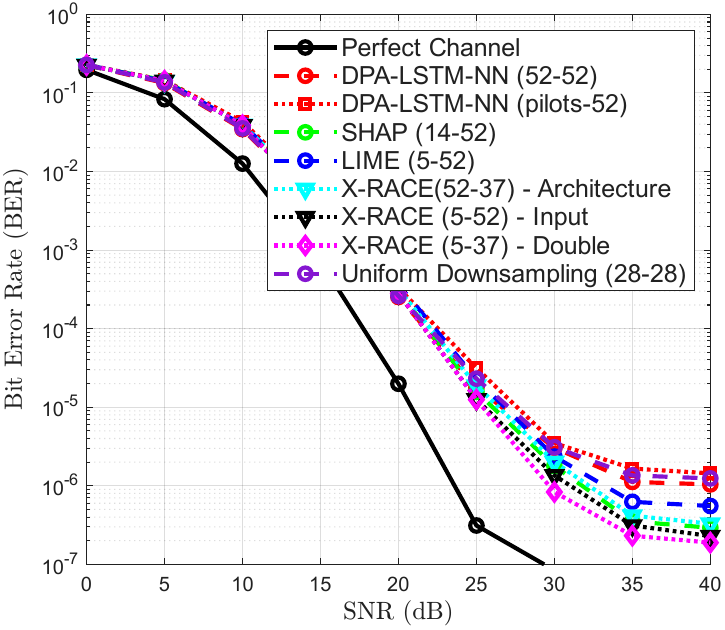}}%
    \hfill
    \subfloat[HF - QPSK.]{\includegraphics[width=0.25\textwidth]{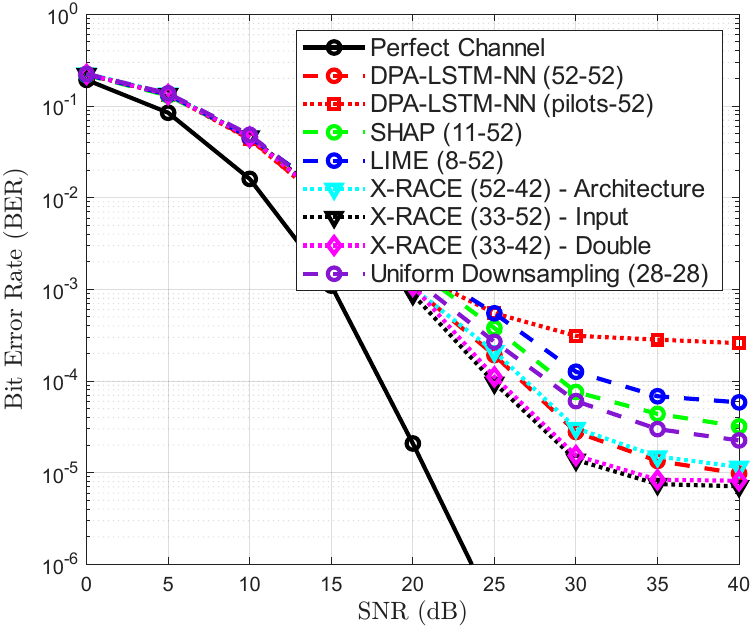}}%
    \hfill
    \subfloat[LF - 64QAM.]{\includegraphics[width=0.25\textwidth]{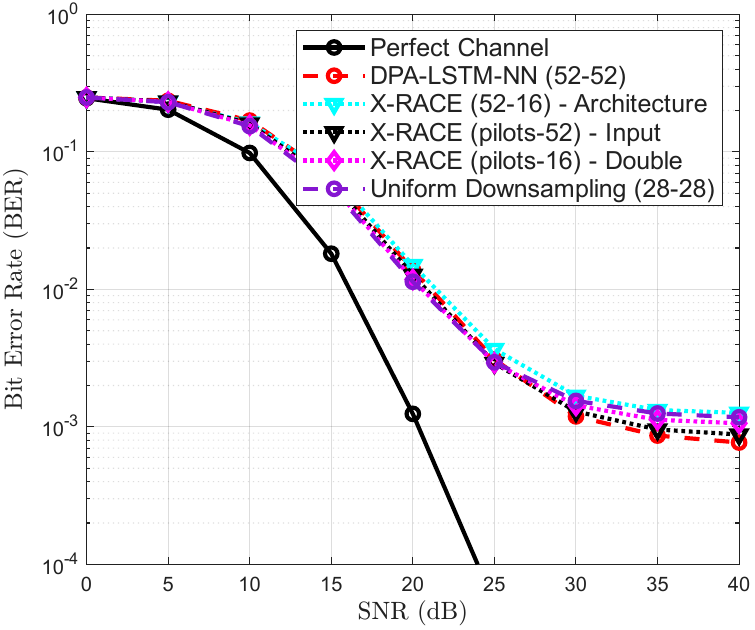}} 
    \hfill
    \subfloat[HF - 64QAM.]{\includegraphics[width=0.25\textwidth]{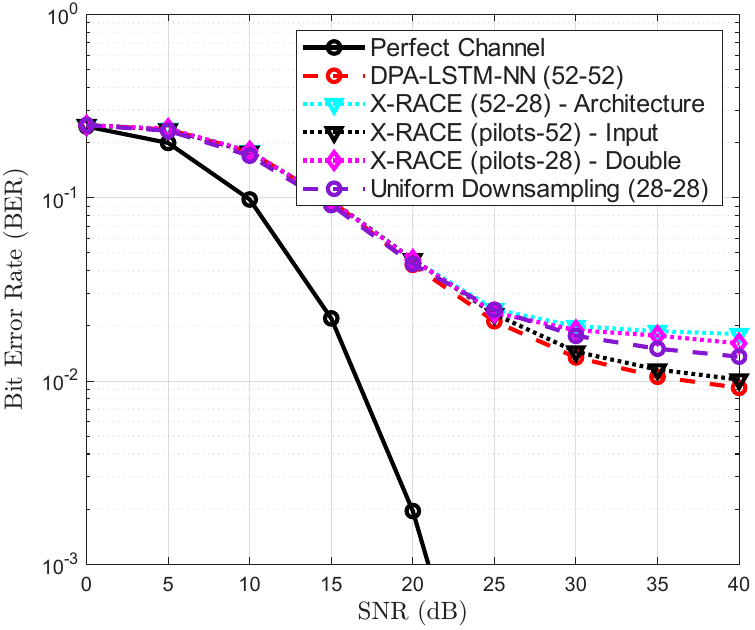}}
    \caption{Relevance score distributions (top) and BER performance (bottom) of the proposed X-RACE framework and benchmarked XAI schemes. For all evaluated schemes, the BER curves correspond to the minimum achieved BER with respect to the best relevance threshold.}
    \label{fig:BER}
\end{figure*}

\subsection{Impact of Frequency Selectivity}

The impact of frequency selectivity is studied under QPSK and $F_d = 1000\text{ Hz}$. As shown in Figures 3(a)-(b), X-RACE consistently assigns pilots the highest relevance score of $0.9$, in both LF and HF scenarios. This proves that pilot relevance is independent of frequency selectivity. However, in the HF scenario, X-RACE assigns fewer data subcarriers a neutral zero score, indicating that increased frequency selectivity demands more informative data subcarriers. Regarding BER performance, X-RACE's double optimization provides BER performance superiority in the LF scenario using only $5$ inputs (threshold $0.3$) and $37$ internal hidden units, as shown in Figure 3(e). This outperforms SHAP, LIME, and uniform downsampling, which rely on larger relevant subsets without yielding BER performance gains. Conversely, in the HF scenario, X-RACE dynamically scales to $33$ relevant inputs (relevance threshold $0.1$) and $42$ internal hidden units, confirming that severe frequency selectivity necessitates increased relevant subcarriers and model capacity.

\subsection{Impact of Modulation Order}

Figures 3(c)-(d) show that under 64QAM, all XAI methods select pilots as the only relevant inputs. However, X-RACE uniquely maintains uniform, stable relevance scores, whereas LIME and SHAP continue to exhibit score variance. BER performance shows that employing only pilot inputs while retaining the full model architecture successfully matches the unpruned baseline. Conversely, applying architecture or double optimization under 64QAM degrades the BER performance, mainly in the HF scenario as shown in Figure 3(h). Consequently, higher modulation orders allow for aggressive input pruning, but strictly require preserving the original LSTM capacity to preserve the BER performance.

\subsection{Computational Complexity Analysis}

X-RACE delivers a dual computational advantage by minimizing relevance score computation overhead and significantly reducing optimized model complexity. Unlike LIME and SHAP which require expensive iterative perturbations with per-symbol complexities of $\mathcal{O}(D_{\text{LIME}}K_{\text{on}}^2)$ and $\mathcal{O}(D_{\text{SHAP}}K_{\text{on}})$, respectively, X-RACE derives input and memory relevance masks in a single pass with $\mathcal{O}(K_{\text{on}} + S)$ complexity. Furthermore, as summarized in Table I, X-RACE achieves the best performance-complexity-interpretability trade-off. Compared to the full baseline, it reduces the FLOPs by $78.5\%$ (LF-QPSK) while improving BER, and achieves a $61.5\%$ reduction for 64QAM while preserving BER. Unlike alternative schemes in which complexity reduction degrades BER performance, X-RACE demonstrates a superior interpretability resolution.

\begin{table}[t]
\renewcommand{\arraystretch}{1.19}
\centering
\caption{XAI-assisted architectural complexity reduction in terms of Floating-Point Operations (FLOPS).}
\begin{tabular}{|c|c|c|c|c|}
\hline
\begin{tabular}[c]{@{}c@{}}XAI\\ Optimization\end{tabular}        & Scenario & \begin{tabular}[c]{@{}c@{}}Architecture\\ ($K_{\text{on}}$ - S)\end{tabular} & FLOPs & Reduction \\ \hline
Full                    & All        & (52-52)         &   64,896    & *         \\ \hline
\begin{tabular}[c]{@{}c@{}}Uniform\\ Downsampling\end{tabular}     & All        & (28-28)     & 18,816      &     71\%      \\ \hline
\multirow{2}{*}{X-RACE} & LF-QPSK  & (5-37)       &  13,912     & 78.5\%          \\ \cline{2-5}
     & HF-QPSK  & (33-42)    & 36,288      &    44.1\%       \\ \cline{2-5}  \hline
\multirow{2}{*}{SHAP}   & LF-QPSK  & (14-52)              &     33,280  &     48.7\%      \\ \cline{2-5} 
                        & HF-QPSK  & (11-52)             & 30,784      &  52.6\%         \\ \hline
\multirow{2}{*}{LIME}   & LF-QPSK  & (5-52)                                                                   &   25,792    &    60.3\%       \\ \cline{2-5} 
                        & HF-QPSK  & (8-52)       &    28,288   &  56.4\%         \\ \hline
\end{tabular}
\end{table}
\section{Conclusion} \label{conclusions}

This paper introduced X-RACE, a dual-optimization attribution framework that resolves LSTM opacity and overhead in doubly-selective channels by simultaneously pruning input subcarriers and internal hidden units. Furthermore, novel temporal XAI metrics uncovered the LSTM's learning dynamics. Simulations confirm that X-RACE reduces inference complexity by at least $44.1\%$ while improving BER performance, achieving a superior performance-complexity-interpretability trade-off over classical XAI. Future work will extend the framework to incorporate gradient-based XAI and optimize other RNN-based channel estimators.

{\small
\bibliographystyle{IEEEtran}
\bibliography{ref}

@ARTICLE{10929033,
  author={Saad, Walid and Hashash, Omar and Thomas, Christo Kurisummoottil and Chaccour, Christina and Debbah, Mérouane and Mandayam, Narayan and Han, Zhu},
  journal={Proceedings of the IEEE}, 
  title={{Artificial General Intelligence (AGI)-Native Wireless Systems: A Journey Beyond 6G}}, 
  year={2025},
  volume={113},
  number={9},
  pages={849-887},
  doi={10.1109/JPROC.2025.3526887}}

@ARTICLE{11063291,
  author={Majumdar, Sayantini and Wei, Qing and Schwarzmann, Susanna and Trivisonno, Riccardo and Carle, Georg},
  journal={IEEE Communications Standards Magazine}, 
  title={{Toward AI-Native 6G Systems: Standards Enablers for 6G Network Automation}}, 
  year={2026},
  volume={10},
  number={1},
  pages={145-153},
  doi={10.1109/MCOMSTD.2025.3585376}}

@ARTICLE{11585835,
  author={Ferrag, Mohamed Amine and Lakas, Abderrahmane and Debbah, Mérouane},
  journal={IEEE Open Journal of the Communications Society}, 
  title={{6G Needs Agents: Toward Agentic AI-Native Networks for Autonomous Intelligence}}, 
  year={2026},
  volume={7},
  number={},
  pages={7254-7282},
  doi={10.1109/OJCOMS.2026.3707904}}

@ARTICLE{11574706,
  author={Kandali, Khalid and Nouh, Said},
  journal={IEEE Open Journal of Intelligent Transportation Systems}, 
  title={{AI-Native V2X and Internet of Vehicles Systems: Architectures, Learning Paradigms, and Open Challenges}}, 
  year={2026},
  volume={7},
  number={},
  pages={1775-1788},
  doi={10.1109/OJITS.2026.3706124}}

@ARTICLE{9954409,
  author={Dos Reis, Ana Flávia and Medjahdi, Yahia and Chang, Bruno Sens and Sublime, Jérémie and Brante, Glauber and Bader, C. Faouzi},
  journal={IEEE Access}, 
  title={{Low Complexity LSTM-NN-Based Receiver for Vehicular Communications in the Presence of High-Power Amplifier Distortions}}, 
  year={2022},
  volume={10},
  number={},
  pages={121985-122000},
  doi={10.1109/ACCESS.2022.3223113}}

@ARTICLE{10540188,
  author={Reis, Ana Flávia dos and Chang, Bruno Sens and Medjahdi, Yahia and Brante, Glauber and Bader, Faouzi},
  journal={IEEE Transactions on Vehicular Technology}, 
  title={{LSTM-Based Time-Frequency Domain Channel Estimation for OTFS Modulation}}, 
  year={2024},
  volume={73},
  number={10},
  pages={15049-15060},
  doi={10.1109/TVT.2024.3406192}}

@inproceedings{lundberg2017unified,
author = {Lundberg, Scott M. and Lee, Su-In},
title = {{A Unified Approach to Interpreting Model Predictions}},
year = {2017},
isbn = {9781510860964},
publisher = {Curran Associates Inc.},
address = {Red Hook, NY, USA},
booktitle = {Proceedings of the 31st International Conference on Neural Information Processing Systems},
pages = {4768–4777},
numpages = {10},
location = {Long Beach, California, USA},
series = {NIPS'17}
}

@inproceedings{ribeiro2016should,
  title={{Why Should I Trust You? Explaining the Predictions of Any Classifier}},
  author={Ribeiro, Marco Tulio and Singh, Sameer and Guestrin, Carlos},
  booktitle={Proceedings of the 22nd ACM SIGKDD international conference on knowledge discovery and data mining},
  pages={1135--1144},
  year={2016}
}

@ARTICLE{10620685,
  author={Senevirathna, Thulitha and La, Vinh Hoa and Marcha, Samuel and Siniarski, Bartlomiej and Liyanage, Madhusanka and Wang, Shen},
  journal={IEEE Communications Surveys \& Tutorials}, 
  title={{A Survey on XAI for 5G and Beyond Security: Technical Aspects, Challenges and Research Directions}}, 
  year={2025},
  volume={27},
  number={2},
  pages={941-973},
  doi={10.1109/COMST.2024.3437248}}

@article{gizzini2025explainable,
  title={{Explainable AI for Enhancing Efficiency of DL-based Channel Estimation}},
  author={Gizzini, Abdul Karim and Medjahdi, Yahia and Ghandour, Ali J and Clavier, Laurent},
  journal={IEEE Transactions on Machine Learning in Communications and Networking},
  year={2025},
  publisher={IEEE}
}

@ARTICLE{10854503,
  author={Sun, Haochen and Liu, Yifan and Al-Tahmeesschi, Ahmed and Nag, Avishek and Soleimanpour, Mohadeseh and Canberk, Berk and Arslan, Hüseyin and Ahmadi, Hamed},
  journal={IEEE Open Journal of the Communications Society}, 
  title={{Advancing 6G: Survey for Explainable AI on Communications and Network Slicing}}, 
  year={2025},
  volume={6},
  number={},
  pages={1372-1412},
  doi={10.1109/OJCOMS.2025.3534626}}

@ARTICLE{r19,
  author={I. {Sen} and D. W. {Matolak}},
  journal={IEEE Transactions on Intelligent Transportation Systems}, 
  title={{Vehicle–Vehicle Channel Models for the 5-GHz Band}}, 
  year={2008},
  volume={9},
  number={2},
  pages={235-245},}
}

\end{document}